\documentclass[12pt]{article}
\usepackage{amsmath}
\usepackage{graphics}

\begin{document}

\vskip 2cm
\title{Open Strings in the SL(2,R) WZWN Model with Solution for a Rigidly
Rotating String}
\author{\\
M.A. Lomholt${}^{*}$
and
A.L. Larsen${}^{|}$}
\maketitle
\noindent
{\em Physics Department, University of Odense, Campusvej 55, 5230
Odense M,
Denmark}

\vskip 8cm
\noindent
$^{*}$Electronic address: mlomholt@fysik.sdu.dk\\
$^{|}$Electronic address: all@fysik.sdu.dk

\newpage
\begin{abstract}
\baselineskip=1.5em
\hspace*{-6mm}Boundary conditions and gluing conditions for open strings
and
D-branes in the
$SL(2,R)$ WZWN model, corresponding to $AdS_3$, are discussed. Some
boundary
conditions and gluing conditions previously considered in the literature
are shown
to be incompatible with the variation principle.

We then consider open string boundary conditions corresponding to a
certain
{\it field-dependent} gluing condition. This allows us to consider open
strings with
constant energy and angular momentum. Classically, these open strings
naturally
generalize the open strings in flat Minkowski space. For rigidly rotating
open
strings, we show that the  torsion leads to a bending and an unfolding. We
also
derive the
$SL(2,R)$ Regge relation, which generalizes the linear Minkowski Regge
relation.
For "high" mass, it takes the form $L\approx \pm M/H$, where $H$ is the
scale of
the $SL(2,R)$ group manifold.
\end{abstract}

\newpage
\section{Introduction}
\label{intro}
\setcounter{equation}{0}
Historically \cite{nambu1,nielsen,susskind}, one of the reasons for string
theory to enter the field of theoretical high energy physics, was its
ability to
reproduce the Regge behaviour seen in the hadron spectrum: As is now well
known,
using the Nambu-Goto action \cite{nambu2,goto} in flat Minkowski space and
imposing standard Neumann boundary conditions one obtains, for a rigidly
rotating open string, the relation
$L=M^2\alpha'$, where $M$ is the mass, $L$ is the angular momentum and
$\alpha'$ is the reciprocal string tension (in suitable units). Rigidly
rotating open strings have been considered also in various black hole and
cosmological backgrounds, and their physical properties have been analyzed
in some detail \cite{inigo,frolov,kar}.

In superstring theory it was for many years the closed strings that
attracted most of the attention (for a review, see for instance
\cite{green}). However, open strings dramatically re-entered the scene due
to the work by Polchinski on D-branes \cite{pol1} (for a review, see for
instance \cite{pol2,pol3}). Open strings and D-branes on group manifolds
have
attracted a lot of interest (see for instance
\cite{klimcik,kato,tseytlin,alekseev}), and recently, especially in
connection with $2+1$-dimensional Anti de Sitter space
\cite{stanciu,bachas,ribault,lee,bachas2} (and references given therein).

$AdS_3$ and the corresponding $SL(2,R)$ WZWN model \cite{witten} plays an
important role in string theory, since it represents a non-trivial (with
curved space and curved time) exact string background (some of the
original works include [23-32]). Moreover, Anti de Sitter space appears in
connection with the  Maldacena conjecture \cite{malda} relating
supergravity and superstring theory with a conformal field theory on the
boundary. In such constructions, $AdS_3$ often appears on the
10-dimensional supergravity/superstring side in a product with some
compact spaces, for instance as $AdS_3\times S^3\times T^4$.

The classical WZWN action \cite{witten} is somewhat ambiguous for an open
string. Basically the problem is that the variation of the action does not
uniquely specify the open string surface terms (boundary terms).
Adding different open string surface terms, which corresponds to a coupling of
the string endpoints to different background vector fields, thus defines
different open string theories. The open string surface terms, in turn,
must
be canceled by imposing appropriate open string boundary conditions;
Neumann, Dirichlet or combinations or generalizations thereof.

Usually D-branes, on which open strings can end, in WZWN models are
described in
terms of gluing conditions. A key-problem concerns the relation between
boundary
conditions and gluing conditions
\cite{klimcik,kato,tseytlin,alekseev,stanciu}.

In the present paper we show that the previously considered Neumann
boundary
condition $\partial_\sigma g=0$ \cite{bars1,bars2} and Neumann type gluing
condition $J=\bar{J}$ for open strings in $AdS_3$ are
incompatible with the variation principle. More precisely, it is
impossible to add an open string surface term to the variation of the
action, which vanishes for such boundary/gluing conditions. The Dirichlet
type gluing condition $J=-\bar{J}$, on the other hand, is
shown to be compatible with the variation principle, provided that the
string endpoints are fixed to move on certain D-branes. These D-branes,
however, are unphysical \cite{bachas}. Eventually, by using another set of
gluing conditions at the string endpoints \cite{stanciu,bachas}, it is
possible to describe open strings attached to physically well-defined
D-branes, and at the same time being compatible with the variation
principle. The latter case is of particular importance since these gluing
conditions preserve the spectral flow \cite{maans2,ooguri}, and the open
strings
can be consistently quantized \cite{ribault,lee} by generalizing the
procedure of \cite{ooguri}.

Finally, we consider an alternative open string surface term as well as
its corresponding boundary conditions. They correspond to a certain
{\it field-dependent} gluing condition. This allows us to consider open strings
with
constant energy and angular momentum. These open strings naturally
generalize the
open strings in flat Minkowski space. For rigidly rotating open strings we
show
that the  torsion leads to a bending and an unfolding of the strings. We
also derive
the
$SL(2,R)$ Regge relation, which generalizes the linear Minkowski Regge
relation.

The paper is organized as follows. In Section 2, we set our notation and
conventions, and we give a general discussion of open string boundary
conditions
and gluing conditions in the $SL(2,R)$ WZWN model. In Section 3, we
consider
rigidly rotating open strings corresponding to a  field-dependent gluing
condition.
The dynamics is solved explicitly and discussed in detail and compared
with the
Minkowski case. In Section 4, we give our conclusions.
\section{The WZWN Action and Open Strings}
\setcounter{equation}{0}

Our starting point is the action for the WZWN model
\cite{witten}\footnote{Our conventions are $\eta^{00}=-1$, $\eta^{11}=1$
and $\epsilon^{01}=1$.}
\begin{eqnarray}\label{eq:WZW}
S^{\rm Closed}&=&S_1+S_2^{\rm Closed}\nonumber\\
S_1&=&-\frac{k}{8\pi}\int_M d\tau d\sigma\ \eta^{\alpha\beta}
\textrm{Tr}\left[g^{-1}\partial_\alpha g\ g^{-1}\partial_\beta
g\right]\nonumber\\
S_2^{\rm Closed}&=&\frac{k}{12\pi}\int_B \textrm{Tr}\left[g^{-1}dg \wedge
g^{-1}dg \wedge g^{-1}dg\right] \label{eq:action2}
\end{eqnarray}
Here $M$ is the string world-sheet and $B$ is a manifold which has $M$ as
its boundary. As the superscript indicates, the WZWN term $S_2^{\rm
Closed}$ is only
defined for closed strings since a boundary cannot have a boundary. Hence,
$M$ cannot be an open string world-sheet. But if we take the variation of
$S_2^{\rm Closed}$ with respect to the group element $g$
\begin{equation}
\delta S_2^{\rm Closed}=- \frac{k}{4\pi}\int_B
d\textrm{Tr}\left[d(g^{-1}dg)\ g^{-1}\delta g\right]
\end{equation}
and use Stokes theorem $\int_B d\omega=\int_M \omega$ for a two-form
$\omega$, with the convention $\int_M F\ d\tau\wedge d\sigma = \int_M
Fd\tau d\sigma$ for a function $F$, we end up with
\begin{equation}\label{eq:WZWvar}
\delta S_2^{\rm Closed}=-\frac{k}{4\pi}\int_M d\tau d\sigma\
\epsilon^{\alpha\beta}\textrm{Tr}\left[\partial_\alpha
(g^{-1}\partial_\beta g)g^{-1}\delta g\right]
\end{equation}
This formula does not refer to the manifold $B$, so one approach for
making a model for open strings is to use this as a starting point. The
problem is that we derived it for closed strings and therefore do not know
anything about a possible surface term.

The $S_1$ term makes perfectly sense for open strings, and taking the
variation of it we get
\begin{eqnarray}
\delta S_1&=&\frac{k}{4\pi}\int_M d\tau d\sigma\
\eta^{\alpha\beta}\textrm{Tr}\left[\partial_\alpha (g^{-1}\partial_\beta
g)g^{-1}\delta g\right]
\nonumber\\
  & & -\frac{k}{4\pi}\int d\tau\ \textrm{Tr}\left[g^{-1}\partial_\sigma
g\,g^{-1}\delta g\right]|_{\sigma=0}^{\sigma=\pi}\
\end{eqnarray}
Adding this to eq.(\ref{eq:WZWvar}), introducing world-sheet light-cone
coordinates $\sigma^\pm=\tau\pm\sigma$ and  the unknown surface term
$\delta S^{\rm Surface}_2$, we have
\begin{eqnarray}
\delta S^{\rm Open}&=&-\frac{k}{2\pi}\int_M d\sigma^- d\sigma^+\
\textrm{Tr}\left[ \partial_- (g^{-1}\partial_+ g)g^{-1}\delta g\right]
\nonumber\\
  & & -\frac{k}{4\pi}\int d\tau\ \textrm{Tr}\left[g^{-1}\partial_\sigma
g\,g^{-1}\delta g\right]|_{\sigma=0}^{\sigma=\pi}+\delta S^{\rm Surface}_2
\end{eqnarray}
The surface terms do not contribute to the equations of motion, so they
are the same for both open and closed strings, namely
\begin{equation}\label{eq:eom}
  \partial_- (g^{-1}\partial_+ g)=0
\end{equation}
which is equivalent to
\begin{equation}
  \partial_+ (\partial_- g\ g^{-1})=0\
\end{equation}
We can therefore define the two quantities
\begin{equation}\label{eq:Jquan}
J=\frac{ik}{2}\partial_- g\ g^{-1}\ ,\quad \bar{J}=\frac{ik}{2}
g^{-1}\partial_+ g\
\end{equation}
depending only on $\sigma^-$ and $\sigma^+$, respectively. From now on
concentrating on the $SL(2,R)$ case, the currents $J$ and $\bar{J}$ take
values in the Lie algebra of $SL(2,R)$. They can be decomposed as
\begin{equation}\label{eq:decomp}
J=\eta_{ab}J^at^b\ ,\quad \bar{J}=\eta_{ab}\bar{J}^at^b
\end{equation}
where $J^a$, $\bar{J}^a$ are real valued fields,
$\eta_{ab}=\textrm{diag}(1,1,-1)$ and the $t^a$ are the three generators
of the $SL(2,R)$ Lie algebra
\begin{equation}
t^1=-\frac{i}{2}\left(\begin{array}{cc}
0&-1\\
-1&0
\end{array}\right)\ ,\quad
t^2=-\frac{i}{2}\left(\begin{array}{cc}
1&0\\
0&-1
\end{array}\right)\ ,\quad
t^3=-\frac{i}{2}\left(\begin{array}{cc}
0&1\\
-1&0
\end{array}\right)
\end{equation}

The currents $J$ and $\bar{J}$ are conserved, as follows from the
equations of motion, but for an open string they can generally not be
obtained as N\"{o}ther currents corresponding to some global symmetries.
Notice also that for an open string, $J$ and $\bar{J}$ are not
independent, since they will in general be related by the boundary
conditions. Moreover, for an open string the corresponding "charges"
\begin{equation}
Q^a_\pm \propto \int_0^\pi d\sigma (J^a \pm \bar{J}^a)
\end{equation}
will in general not be constants of motion. Different open string surface
terms will therefore in general define different open string theories with
different constants of motion.

The equations of motion are as usual supplemented by the constraints
\begin{equation}
\mbox{Tr}[g^{-1}\partial_\pm g\  g^{-1}\partial_\pm g]=0
\end{equation}
which correspond to vanishing world-sheet energy-momentum tensor.

\subsection{The Surface Term}
All we need to do now is to specify the surface term $\delta S^{\rm
Surface}_2$. In Refs.\cite{bars1,bars2} it was set equal to zero, but if
we want $\delta S^{\rm Open}$ to be the variation of some action $S^{\rm
Open}$, we cannot choose $\delta S^{\rm Surface}_2$ freely, and especially
setting it to zero will not work. To see this for the case of $SL(2,R)$,
we parametrise the group elements as
\begin{equation}
g= e^{2uit^3}e^{2\rho it^2}e^{2vit^3}
\end{equation}
where $u$, $\rho$ and $v$ are the new fields on the string world-sheet,
and the $t^a$ are the generators of the $SL(2,R)$ Lie algebra introduced
before. If $\delta S^{\rm Open}$ is the variation of some action $S^{\rm
Open}$, then we should be able to take the variation of $\delta S^{\rm
Open}$ and get something that is symmetric in the variations of the
fields. That is, if we take the variation twice, with respect to for
instance $\rho$ and $u$, we should get the same result independently of
the order in which we take the variations. So we have the condition
\begin{equation}\label{eq:cond}
\delta_{{\delta X}^\mu}\delta_{{\delta X}^\nu}S^{\rm Open}=\delta_{{\delta
X}^\nu}\delta_{{\delta X}^\mu}S^{\rm Open}
\end{equation}
where $X^\mu$ and $X^\nu$ are any of the fields $u$, $\rho$ or $v$.
Concentrating on the WZWN term, where the problems might arise, we can write
the variation as
\begin{equation}
\delta S_2^{\rm Open}=\delta S_2^{\rm Bulk}+\delta S_2^{\rm Surface}
\end{equation}
where $\delta S_2^{\rm Bulk}$ has the same expression as $\delta S_2^{\rm
Closed}$ in eq.(\ref{eq:WZWvar}). If we set $\delta S_2^{\rm Surface}$
equal to zero (as in Refs.\cite{bars1,bars2}), we only have $\delta
S_2^{\rm Bulk}$ left. Inserting the $SL(2,R)$ parametrisation in
eq.(\ref{eq:WZWvar}), we get
\begin{eqnarray}
\delta S_2^{\rm Bulk}=\frac{k}{\pi}\int_M d\tau d\sigma\ \sinh 2\rho\
\epsilon^{\alpha\beta}[ \partial_\alpha v\partial_\beta\rho\,\delta u +
\partial_\alpha\rho\partial_\beta u\,\delta v + \partial_\alpha
u\partial_\beta v\,\delta\rho]\
\end{eqnarray}
It follows that
\begin{eqnarray}
\delta_{\delta \rho} S_2^{\rm Bulk}&=& \frac{k}{\pi}\int_M d\tau d\sigma\
\epsilon^{\alpha\beta}\sinh 2\rho\ \partial_\alpha u\partial_\beta
v\,\delta\rho\ \\
\delta_{\delta u} S_2^{\rm Bulk}&=& \frac{k}{\pi}\int_M d\tau d\sigma\
\epsilon^{\alpha\beta}\sinh 2\rho\ \partial_\alpha
v\partial_\beta\rho\,\delta u\
\end{eqnarray}
After another variation we get
\begin{eqnarray}
\delta_{\delta u}\delta_{\delta\rho}S_2^{\rm Bulk}&=&-\frac{k}{\pi}\int_M
d\tau d\sigma\ \epsilon^{\alpha\beta}\sinh 2\rho\ \partial_\alpha
v\partial_\beta\delta u\,\delta\rho\ \\
\delta_{\delta \rho}\delta_{\delta u}S_2^{\rm Bulk}&=&-\frac{k}{\pi}\int_M
d\tau d\sigma\ \epsilon^{\alpha\beta}\sinh 2\rho\ \partial_\alpha
v\partial_\beta\delta u\,\delta\rho \nonumber \\
&&+\frac{k}{\pi}\int d\tau\ \sinh 2\rho\ \partial_\tau v\ \delta\rho\delta
u|_{\sigma=0}^{\sigma=\pi}
\end{eqnarray}
We see that the condition (\ref{eq:cond}) is not fulfilled, and hence
there is no $S^{\rm Open}$ with  vanishing $\delta S_2^{\rm Surface}$. If
we want the open string theory to be well defined in terms of an action,
we therefore cannot proceed as in Refs.\cite{bars1,bars2} and forget about
a surface term from the WZWN term.

\subsection{An Open String Action}
A way to avoid the above problems is to introduce a parametrisation in
eq.(\ref{eq:action2}), use Stokes theorem to write it as an integral over
$M$, and then use that expression to define an action for open strings.
Inserting the $SL(2,R)$ parametrisation in eq.(\ref{eq:action2}), we find
\begin{eqnarray}
S_2^{\rm Closed}&=& \frac{k}{\pi}\int_B \sinh 2\rho\ d\rho\wedge du\wedge
dv \nonumber\\
&=& \frac{k}{\pi}\int_B d\left(\sinh^2\rho\ du\wedge dv\right) \nonumber\\
&=&\frac{k}{\pi}\int_M d\tau d\sigma\ \epsilon^{\alpha\beta}\sinh^2\rho\
\partial_\alpha u\partial_\beta v
\end{eqnarray}
The last expression can then be used to define the open string action
\begin{equation}\label{eq:openaction2}
S_2^{\rm Open}=\frac{k}{\pi}\int_M d\tau d\sigma\
\epsilon^{\alpha\beta}\sinh^2\rho\ \partial_\alpha u\partial_\beta v
\end{equation}
It is important to stress that there is some ambiguity in the above
procedure. The action for the closed string is the same if we add a total
$\sigma$ derivative to the integrand of the action. This is not true for
the open string action. Said in another way, there are in some sense many
open string actions corresponding to a single closed string action. But
the special choice (\ref{eq:openaction2}) has some nice features.

If we introduce a new parametrisation ($H$ is a scale of the $SL(2,R)$
group-manifold)
\begin{equation}
\sinh \rho=Hr\ ,\quad u=\frac{1}{2}(Ht+\phi)\ ,\quad
v=\frac{1}{2}(Ht-\phi)
\end{equation}
in terms of which
\begin{equation}
g= \left(\begin{array}{cc}
\sqrt{1+H^2r^2}\cos Ht+Hr\cos\phi&\sqrt{1+H^2r^2}\sin Ht-Hr\sin\phi\\
-\sqrt{1+H^2r^2}\sin Ht-Hr\sin\phi&\sqrt{1+H^2r^2}\cos Ht-Hr\cos\phi
\end{array}\right)\
\end{equation}
we get the following total action (well defined for both closed and open
strings)
\begin{eqnarray}
S&=-\frac{H^2k}{4\pi}\int_M d\tau d\sigma\
\Big\{&\eta^{\alpha\beta}\left[-(1+H^2r^2) \partial_\alpha t\partial_\beta
t+\frac{\partial_\alpha r\partial_\beta r}{1+H^2r^2}+r^2\partial_\alpha
\phi\partial_\beta \phi\right]\nonumber\\
&&+2\epsilon^{\alpha\beta}Hr^2\partial_\alpha t\partial_\beta \phi\ \
\Big\} \label{eq:adsaction}
\end{eqnarray}
If we let $H\to 0$, scaling $k$ such that $1/\alpha' =H^2k$ is kept
constant, (\ref{eq:adsaction}) becomes the usual Minkowski action in polar
coordinates. Also a nice feature is that (\ref{eq:adsaction}) is invariant
under global $t$ and $\phi$ translations, so we can define constant energy
and angular momentum as the corresponding N\"{o}ther charges.

For $SL(2,R)$ we get the same group element $g$ if we translate $t$ by
$2\pi$. We will unwrap the time coordinate so that $t$ is not identified
with $t+2\pi$. As usual, this corresponds to going to the universal cover
of $SL(2,R)$.

\subsection{Boundary Conditions}
The variation of (\ref{eq:adsaction}) is (a prime/dot denotes a
$\sigma$/$\tau$ derivative)
\begin{eqnarray}
\delta S=&-&\frac{H^2k}{2\pi}\int_M d\tau d\sigma\ \Big\{
\Big[-\frac{r''-\ddot r}{1+H^2r^2}-H^2r(t't'-\dot t\dot t)+
\frac{H^2r(r'r'-\dot r\dot r)}{(1+H^2r^2)^2}\nonumber\\
&\ &\ \ +r(\phi'\phi'-\dot\phi\dot\phi)+2Hr(\phi'\dot
t-t'\dot\phi)\Big]\delta r\nonumber\\
&\ &\ \ +\Big[(1+H^2r^2)(t''-\ddot t)+2H^2r(t'r'-\dot t\dot
r)+2Hr(r'\dot\phi-\phi'\dot r) \Big]\delta t\nonumber\\
&\ &\ \ +\Big[-r^2(\phi''-\ddot\phi)-2r(r'\phi'-\dot r\dot\phi)+2Hr(t'\dot
r-r'\dot t)\Big]\delta\phi\ \Big\}
\nonumber\\
&-&\frac{H^2k}{2\pi}\int d\tau\ \Big\{
\Big[-(1+H^2r^2)t'-Hr^2\dot\phi\Big]\delta
t+\Big[\frac{r'}{1+H^2r^2}\Big]\delta r\nonumber\\
&\ &\ \ +\Big[r^2\phi'+Hr^2\dot t\Big]\delta\phi\
\Big\}\Big|_{\sigma=0}^{\sigma=\pi} \label{eq:varadsact}
\end{eqnarray}
so, besides the equations of motion, we get the boundary conditions
\begin{eqnarray}
t'=&-\frac{Hr^2}{(1+H^2r^2)}\dot\phi &;\ \sigma=0,\pi \nonumber\\
r'=&0 &;\ \sigma=0,\pi \\
\phi'=&-H\dot t &;\ \sigma=0,\pi \nonumber
\end{eqnarray}

As remarked earlier, we can add to the action a total sigma derivative
\begin{equation}
S_3=-\int_M d\tau d\sigma\ \partial_\sigma A=-\int d\tau\ A\
|_{\sigma=0}^{\sigma=\pi}
\end{equation}
without affecting the equations of motion. If we want the boundary
conditions to be linear in derivatives, then $A$ must have the form
$A=A_\mu\partial_\tau X^\mu$ (there cannot be any $\sigma$ derivatives if
the Lagrange density does not contain any derivatives of higher order than
one). If we introduce $F_{\mu\nu}\equiv A_{\nu,\mu}-A_{\mu,\nu}$ we can
write $S_3$ in two equivalent ways
\begin{equation}
S_3=\int_M d\tau d\sigma\ F_{\mu\nu}\partial_\tau X^\mu\partial_\sigma
X^\nu= -\int d\tau\ A_\mu\partial_\tau X^\mu\ |_{\sigma=0}^{\sigma=\pi}
\end{equation}
and we see that this term can be interpreted in two ways. Either as an
addition of a total derivative to the antisymmetric tensor background, or
as a coupling of the string endpoints to a vector field background.

For the variation of the action we get the extra term
\begin{equation}\label{eq:varadsact3}
\delta S_3=\int d\tau\ F_{\mu\nu}\partial_\tau X^\mu\delta X^\nu
|_{\sigma=0}^{\sigma=\pi}
\end{equation}
From this expression and eq.(\ref{eq:varadsact}), we see that we now have the
more general set of boundary conditions at $\sigma=0,\pi$
\begin{eqnarray}
(1+H^2r^2)t'&=&-Hr^2\dot\phi -{\tilde F}_{\mu t}\partial_\tau X^\mu
\nonumber \\
\frac{r'}{1+H^2r^2}&=&{\tilde F}_{\mu r}\partial_\tau X^\mu
\label{eq:bound_cond}\\
r^2\phi'&=&-Hr^2\dot t +{\tilde F}_{\mu\phi}\partial_\tau X^\mu \nonumber
\end{eqnarray}
where ${\tilde F}_{\mu\nu}\equiv \frac{2\pi}{H^2k}F_{\mu\nu}$.

Now a question one can ask is the following: Given some boundary
conditions, is it possible to find $A_\mu$ such that the boundary
conditions can be derived from the action (\ref{eq:adsaction}) with $S_3$
added to it? If this is possible, we should be able to write the boundary
conditions with all the $\sigma$ derivatives on the left hand side of the
equality sign, in the same form as above, and read off the tensor ${\tilde
F}_{\mu\nu}$ from the right hand side. ${\tilde F}_{\mu\nu}$ should then
be antisymmetric and satisfy the condition (the square of the exterior
derivative being zero)
\begin{equation}
{\tilde F}_{\mu\nu,\rho}+{\tilde F}_{\rho\mu,\nu}+{\tilde
F}_{\nu\rho,\mu}=0
\end{equation}

As a simple example, one can ask if it is possible to choose $A_\mu$ such
that there are no $\tau$ derivatives in the boundary conditions
(corresponding to only taking the boundary conditions from the $S_1$ part
of the action). Such boundary conditions were considered in
Refs.\cite{bars1,bars2}. They correspond to the standard Neumann boundary
conditions, which are usually also imposed for open strings in flat
Minkowski space (see for instance \cite{green}). However, for the $\tau$
derivatives to disappear in eq.(2.31), we must have
\begin{equation}
{\tilde F}_{\phi t}=-{\tilde F}_{t\phi}=-Hr^2\ ,\quad {\tilde
F}_{tr}={\tilde F}_{rt}={\tilde F}_{r\phi}={\tilde F}_{\phi r}=0
\end{equation}
This is antisymmetric, but
\begin{equation}
{\tilde F}_{\phi t,r}+{\tilde F}_{r\phi,t}+{\tilde F}_{tr,\phi}=-2Hr
\end{equation}
does not vanish, and so we cannot find a $A_\mu$ giving a boundary
condition without $\tau$ derivatives. Thus it is not possible to obtain
the standard Neumann boundary conditions, from an action principle, in the
case of the WZWN model corresponding to $SL(2,R)$. This is of course
consistent with the conclusion obtained in Subsection 2.1.

\subsection{The Neumann Type Gluing Condition}
A lot of work (see for instance
Refs.\cite{kato,tseytlin,alekseev,stanciu}) on D-branes in the WZWN models
use the gluing conditions (evaluated at $\sigma=0,\pi$. We will not write
this in the rest of this section)
\begin{equation}
J=\pm \bar J
\end{equation}
where $J$ and $\bar J$ are defined in eq.(\ref{eq:Jquan}). One of them is
considered as a generalization of the Neumann boundary conditions
$\partial_\sigma X^\mu=0$ in flat space, and the other as a generalization
of the corresponding Dirichlet boundary conditions $\partial_\tau
X^\mu=0$. To see which is which, consider the case where the
group-manifold is Abelian. If the group elements $g$ are parametrised by
$g=e^{iX}$, with $X$ in the Lie algebra, then
\begin{eqnarray}
J=+\bar J\ \Leftrightarrow&\partial_-X=\partial_+X &\Leftrightarrow\
\partial_\sigma X=0 \nonumber\\
J=-\bar J\ \Leftrightarrow&\partial_-X=-\partial_+X &\Leftrightarrow\
\partial_\tau X=0
\end{eqnarray}
and we see that $J=\bar J$ corresponds to the Neumann conditions.

As another example of the procedure developed in the last subsection, we
can try to see if the Neumann type gluing condition can be derived from an
open string $SL(2,R)$ WZWN action. In terms of the ($t,r,\phi$)
parametrisation, we have for the decomposition (\ref{eq:decomp})
\begin{eqnarray}
J^1+iJ^2&=&k\left[Hr\sqrt{1+H^2r^2}\left(H\partial_-t-\partial_-\phi\right)-
i\frac{H\partial_-r}{\sqrt{1+H^2r^2}}\right]e^{-i(Ht+\phi)}\
\nonumber \\
J^3&=&k\left[(1+H^2r^2)H\partial_-t-H^2r^2\partial_-\phi\right] \nonumber
\\
\bar{J}^1+i\bar{J}^2&=&k\left[Hr\sqrt{1+H^2r^2}
\left(-H\partial_+t-\partial_+\phi\right)-i\frac{H\partial_+r}{\sqrt{1+H^2r^2}}
 \right]e^{i(Ht-\phi)}
\nonumber\\
\bar{J}^3&=&k\left[(1+H^2r^2)H\partial_+t+H^2r^2\partial_+\phi\right]
\end{eqnarray}
Thus, the Neumann type gluing conditions $J^3=\bar{J}^3$,
$J^1+iJ^2=\bar{J}^1+i\bar{J}^2$ are equivalent to
\begin{eqnarray}
(1+H^2r^2)t'&=&-Hr^2\dot\phi \nonumber \\
\frac{r'}{1+H^2r^2}&=&-\frac{r\dot\phi}{1+H^2r^2}\tan Ht\\
r^2\phi'&=&-Hr^2\dot t +\frac{r\dot r}{1+H^2r^2}\tan Ht \nonumber
\end{eqnarray}
Reading off the components of ${\tilde F}_{\mu\nu}$ we get
\begin{equation}
{\tilde F}_{\phi r}=-{\tilde F}_{r\phi}=-\frac{r}{1+H^2r^2}\tan Ht\ ,\quad
{\tilde F}_{tr}={\tilde F}_{rt}={\tilde F}_{t\phi}={\tilde F}_{\phi t}=0
\end{equation}
Again, this is antisymmetric but
\begin{equation}
{\tilde F}_{\phi t,r}+{\tilde F}_{r\phi,t}+{\tilde
F}_{tr,\phi}=\frac{Hr}{1+H^2r^2}(1+\tan^2 Ht)
\end{equation}
so we cannot, in the framework considered in this article, find an action
giving the Neumann type gluing conditions.

\subsection{The Dirichlet Type Gluing Condition}
We can also take a closer look at the Dirichlet type gluing conditions
\begin{equation}\label{eq:glu1}
J=-\bar J
\end{equation}
In the $(t,r,\phi)$ parametrisation they are equivalent to the equations
\begin{eqnarray}
(1+H^2r^2)t'-\frac{r'}{Hr}\tan
Ht&=&-Hr^2\dot\phi-\frac{\dot\phi}{H}\nonumber\\
r^2\phi'&=&-Hr^2\dot t-\frac{\dot t}{H}\label{eq:Dirbound}\\
0&=&\frac{H^2r\dot r}{\sqrt{1+H^2r^2}}\cos Ht-\sqrt{1+H^2r^2}\sin Ht\
H\dot t\nonumber
\end{eqnarray}
and we see that we obviously cannot choose ${\tilde F}_{\mu\nu}$, so that
these equations become equivalent to eqs.(\ref{eq:bound_cond}). This is
all right, since we decided to consider them as a generalization of the
Dirichlet boundary conditions, and therefore they are not supposed to be
derivable from an action, with the endpoints of the string being allowed
to move freely. Instead, the endpoints should be allowed only to move on
some submanifold of space-time, that is a D-brane. By integrating the last
equation, we see that we have a candidate in the D-string consisting of
points satisfying
\begin{equation}\label{eq:circularD}
\sqrt{1+H^2r^2}\cos Ht=C
\end{equation}
where $C$ is the integration constant. The rest of the equations in
(\ref{eq:Dirbound}) are then supposed to be derivable from the variation
of the action, eqs.(\ref{eq:varadsact}, \ref{eq:varadsact3}), with the
restriction from eq.(\ref{eq:circularD}) that there should be the
following relation between the variations of the endpoints
\begin{equation}
\frac{H^2r}{\sqrt{1+H^2r^2}}\cos Ht\ \delta r=\sqrt{1+H^2r^2}\sin Ht\
H\delta t
\end{equation}
This is the case if we choose
\begin{equation}
{\tilde F}_{\phi t}=-{\tilde F}_{t\phi}=\frac{1}{H}\ ,\quad {\tilde
F}_{\phi r}={\tilde F}_{r\phi}={\tilde F}_{rt}={\tilde F}_{tr}=0
\end{equation}
and we see that we finally have a total exterior derivative, so the open
string action is well defined. However, it was argued in Ref.\cite{bachas}
that this D-brane is not physical. The dynamics of a D-brane is governed
by the Dirac-Born-Infeld action
\begin{equation}
S_{\rm DBI}=-T_{\rm D}\int d^{p+1}\xi\ \sqrt{-\det\left({\hat
G}_{ab}+{\hat B}_{ab}+2\pi\alpha'{\hat F}_{ab}\right)}
\end{equation}
Here $T_{\rm D}$ is the tension of the D-brane, $\xi^0$,\dots,$\xi^p$ is a
parametrisation of the D-brane world-volume, and a hat denotes the
pullback to the D-brane world-volume. For instance
\begin{equation}
{\hat G}_{ab}=G_{\mu\nu}\frac{\partial X^\mu}{\partial\xi^a}\frac{\partial
X^\nu}{\partial \xi^b}
\end{equation}
The D-brane (\ref{eq:circularD}) solves the equations of motion derived
from the Dirac-Born-Infeld action, but it is unphysical since the action
is imaginary
\begin{equation}
-\det\left({\hat G}_{ab}+{\hat B}_{ab}+2\pi\alpha'{\hat
F}_{ab}\right)=-\frac{C^2}{H^4\cos^4 \xi^0}<0
\end{equation}
when $C\ne 0$ (we have parametrised the world-volume by $\xi^0=Ht$,
$\xi^1=\phi$).

As a last example, we note that the gluing conditions
\cite{stanciu,bachas}
\begin{equation}\label{eq:glu2}
J^3={\bar J}^3\ ,\quad J^1+iJ^2=e^{2i\alpha}\left({\bar J}^1-i{\bar
J}^2\right)
\end{equation}
where $\alpha$ is an arbitrary constant, can be derived in the same way
from eqs.(\ref{eq:varadsact}, \ref{eq:varadsact3}), if the string
endpoints are restricted to D-branes of the form
\begin{equation}\label{eq:stretchedD}
Hr\cos(\phi+\alpha)=C
\end{equation}
where $C$ is again a constant. This time we have to choose
\begin{equation}
{\tilde F}_{\phi t}={\tilde F}_{t\phi}={\tilde F}_{\phi r}={\tilde
F}_{r\phi}={\tilde F}_{rt}={\tilde F}_{tr}=0
\end{equation}
which is clearly a total derivative. The D-brane (\ref{eq:stretchedD})
solves the Dirac-Born-Infeld equations of motion, and is physical since
the
action is real (we have chosen $\xi^0=Ht$,
$\xi^1=\phi+\alpha$)
\begin{equation}
-\det\left({\hat G}_{ab}+{\hat B}_{ab}+2\pi\alpha'{\hat
F}_{ab}\right)=\frac{C^2}{H^4\cos^4 \xi^1}\ge 0
\end{equation}
as it should be.
This D-brane was also studied more extensively in Ref.\cite{bachas}.

Both sets of gluing conditions (\ref{eq:glu1}) and (\ref{eq:glu2}) are
compatible with the spectral flow considered in \cite{maans2,ooguri},
which takes a solution $\tilde g$ of the equations of motion
(\ref{eq:eom})
and generates the new solution
\begin{equation}
g=e^{w_R\sigma^-it^3}{\tilde g}e^{w_L\sigma^+it^3}
\end{equation}
or in terms of the currents (\ref{eq:decomp})
\begin{eqnarray}
J^1+i J^2&=&\left({\tilde J}^1+i{\tilde J}^2\right)e^{-i w_R\sigma^-}\\
J^3&=&{\tilde J}^3+\frac{k}{2}w_R\\
{\bar J}^1+i {\bar J}^2&=&\left({\tilde{\bar J}}^1+i{\tilde{\bar
J}}^2\right)e^{i w_L\sigma^+}\\
{\bar J}^3&=&{\tilde{\bar J}}^3+\frac{k}{2}w_L
\end{eqnarray}
with the restriction for (\ref{eq:glu1}) that $w_R=-w_L=w$ and for
(\ref{eq:glu2}) that $w_R=w_L=w$, $w$ taking integer values. This
observation was used in \cite{ribault,lee} to generalize the quantization
procedure of \cite{ooguri} to the case of open strings ending on
D-branes
of the form (\ref{eq:stretchedD}).

\section{Rigidly Rotating Strings}
\setcounter{equation}{0}
In this section we shall consider rigidly rotating open strings in the
$SL(2,R)$ WZWN model. We choose to work with the open string action
(2.25), which is invariant under global $t$ and $\phi$ translations. This
way, both the energy and angular momentum are ensured to be constants of
motion. This should be contrasted with the cases considered in Section 2.

The open
string boundary conditions are given by eq.(2.27). Using eq.(2.37), it
follows that
they correspond to the following {\it field-dependent} gluing conditions
\begin{equation}\label{eq:glu3}
J^3={\bar J}^3\ ,\quad J^1+iJ^2=e^{-2iHt}\left({\bar J}^1+i{\bar
J}^2\right)
\end{equation}
The rigidly rotating open string ansatz is
\begin{eqnarray}
&t&=\tilde{t}(\sigma )+c_1\tau \nonumber \\
&\phi& =\tilde{\phi }(\sigma )+c_2\tau \nonumber \\
&r&=r(\sigma )
\end{eqnarray}
where $c_1$ and $c_2$ are constants. We take $c_1>0$ to ensure forward
propagation in time $(\dot{t}>0)$, while $c_2$ is arbitrary. The above
ansatz describes the most general, with constant velocity, rotating rigid
string (up to world-sheet coordinate transformations) \cite{frolov}. The
equations of motion and constraints, in this case, reduce to
\begin{eqnarray}
\frac{d\tilde{t}}{d\sigma} &=&\frac{k_1-Hc_2r^2}{1+H^2r^2} \\
\frac{d\tilde{\phi}}{d\sigma} &=&\frac{k_2-Hc_1r^2}{r^2}\\
\left( \frac{dr}{d\sigma}\right)^2 &+& V(r)=0
\end{eqnarray}
where the potential $V(r)$ is given by
\begin{eqnarray}
V(r)= &-& \frac{(H^2k_2^2-k_1^2)(c_1^2+2Hc_1k_2)}{r^2k_2^2} \nonumber \\
        &\cdot & \big( r^2 - \frac{k_2^2}{c_1^2+2Hc_1k_2} \big) \cdot
\big(r^2 + \frac{k_2^2}{H^2k_2^2-k_1^2} \big)
\end{eqnarray}
and the integration constants $(k_1,k_2)$ are constrained by
\begin{equation}
c_1k_1=c_2k_2
\end{equation}
The boundary conditions (2.27) demand
\begin{equation}
k_1=k_2=0,\ \ V(r)_{|_{\sigma=0,\pi}}=0
\end{equation}
The latter condition gives rise to the relation $c_2^2=n^2+H^2c_1^2$,
where $n$ is an arbitrary integer. Then the solution is given by
\begin{eqnarray}
&r&=\frac{c_1}{n}\cos(n\sigma )\\
&t&=c_1\tau \mp \frac{\sqrt{n^2+H^2c_1^2}}{H}\ \sigma \pm
\frac{1}{H}\cot^{-1} \left( \frac{\sqrt{n^2+H^2c_1^2}}{n}\cot (n\sigma)
\right) \\
&\phi&=\pm\sqrt{n^2+H^2c_1^2}\ \tau-Hc_1\sigma
\end{eqnarray}
The upper sign corresponds to positive values of $c_2$, the lower to
negative. Notice that the integer $n$ plays the role of a winding number
or, more precisely, it gives the number of "foldings" of the open string.
Namely, for $H\rightarrow 0$, we get
\begin{equation}
r=\frac{c_1}{n}\cos(n\sigma),\ \ \ \phi=\pm\frac{|n|}{c_1}t; \ \ \
H\rightarrow 0
\end{equation}
which is the standard $n$-folded rigidly rotating straight string in
Minkowski space. For further comparison with the Minkowski case, it is also
convenient to express $r$ and $\phi$ in terms of $t$ and $\sigma$, and
then to consider the string at fixed coordinate time $t$. In particular,
eq.(3.11) leads to
\begin{eqnarray}
\phi (t,\sigma )=&\pm&\frac{\sqrt{n^2+H^2c_1^2}}{c_1}\
t+\frac{n^2}{Hc_1}\sigma\nonumber\\
&-&\frac{\sqrt{n^2+H^2c_1^2}}{Hc_1}\cot^{-1} \left(
\frac{\sqrt{n^2+H^2c_1^2}}{n}\cot (n\sigma) \right)
\end{eqnarray}
and it follows that the torsion leads to a bending and an unfolding of the
string; see Figure 1. This should be contrasted with the case of Minkowski
space and with the case of "pure" anti de Sitter space without torsion
\cite{inigo,kar}.

Invariance of the action (2.25) under constant $t$ and $\phi$ translations
gives rise to conserved N\"{o}ther currents
\begin{equation}
P^\alpha_t=\frac{\partial {\cal L}}{\partial t_{,\alpha}},\ \ \
P^\alpha_\phi=\frac{\partial {\cal L}}{\partial \phi_{,\alpha}}
\end{equation}
The corresponding constant charges are given by
\begin{eqnarray}
Q_t&=&\int_0^\pi d\sigma P^\tau_{t}=-\frac{H^2k}{2\pi}\int_0^\pi d\sigma
\big[(1+H^2r^2)\dot{t}+ Hr^2\phi^\prime \big] \\
Q_\phi&=&\int_0^\pi d\sigma P^\tau_{\phi}=\frac{H^2k}{2\pi}\int_0^\pi
d\sigma \big[ r^2\dot{\phi}+ Hr^2 t^\prime \big]
\end{eqnarray}
The first charge is identified with minus the mass-energy $M$ and the
second one with angular momentum $L$. Using the solution (3.9)-(3.11), we
get
\begin{equation}
M=\frac{c_1}{2\alpha^\prime},\ \ \ L=\pm\frac{1}{2\alpha^\prime H^2}\left(
\sqrt{n^2+ H^2 c_1^2}-n\right)
\end{equation}
where we used again the relation $1/\alpha^\prime =H^2k$. It follows that
we get the relation
\begin{equation}
L=\pm\frac{M^2\alpha^\prime}{|n|}\left[ \frac{\sqrt{n^4+4H^2\alpha^{\prime
2}M^2n^2}-n^2}{2H^2\alpha^{\prime 2}M^2} \right]
\end{equation}
For $H\rightarrow 0$ ($k\rightarrow \infty$), the square bracket goes to
$1$ and we recover, for $n=1$, the famous Minkowski-space Regge behavior
(see for instance Ref.\cite{green}). Thus, the relation (3.18) can be
considered as the $SL(2,R)$ generalization of the Minkowski Regge
behavior. In the present case, the linear relationship between $L$ and
$M^2$ is recovered only for "small" mass, while for "high" mass we get
instead
\begin{equation}
L\approx \pm M/H
\end{equation}
\section{Conclusion}
\setcounter{equation}{0}
It was recently shown that open strings in $AdS_3$ can be quantized
\cite{ribault,lee}, by generalizing the procedure for closed strings
\cite{ooguri}.
Quantization of open strings involves D-branes, on which the open strings
can
end, corresponding to certain gluing conditions \cite{stanciu,bachas}.
Previously in the literature, various other boundary conditions and gluing
conditions have been considered.

We have shown that some of the  previously
considered  boundary
conditions and gluing
conditions for open strings in $AdS_3$ are in fact
incompatible with the variation principle. Other boundary conditions and
gluing
conditions are compatible with the variation principle, and get
interpretations in
terms of D-branes.

We then considered a certain
field-dependent gluing condition, compatible with the variation principle.
The
corresponding open strings seem to give the most natural generalization,
at
least
classically, of the open strings in flat Minkowski space. The open strings
were
analyzed in detail for an ansatz corresponding to rigid rotation. We
showed, in
particular, that the torsion leads to a bending and an unfolding of the
strings.
Finally, we derived the
$SL(2,R)$ Regge relation, i.e. the relation between mass $M$ and angular
momentum
$L$ for rigidly rotating strings in $SL(2,R)\cong AdS_3$. It turned out to
be
quite different from the Minkowski Regge relation; the linear relationship
between
$L$ and
$M^2$ is recovered only for "small" mass, while for "high" mass we found
instead $L\approx \pm M/H$, where $H$ is the scale of
the $SL(2,R)$ group manifold.

\vskip 24pt
\hspace*{-6mm}{\bf Acknowledgements}:\\
We would like to thank I.L. Egusquiza for discussions on the material
presented in
Section 3.

\newpage
\centerline{\Large CAPTIONS FOR FIGURES} \vskip 2cm

\noindent Figure 1: The rigidly rotating string (3.9)-(3.11) for $n=2$ and
$Hc_1=1$. The string is shown at fixed coordinate time $t=0$. The axes
represent
$Hr\cos\phi$ and $Hr\sin\phi$, respectively. Notice that the string is
bended and
unfolded.
\newpage
\begin{figure}[htb]
\begin{center}
\includegraphics{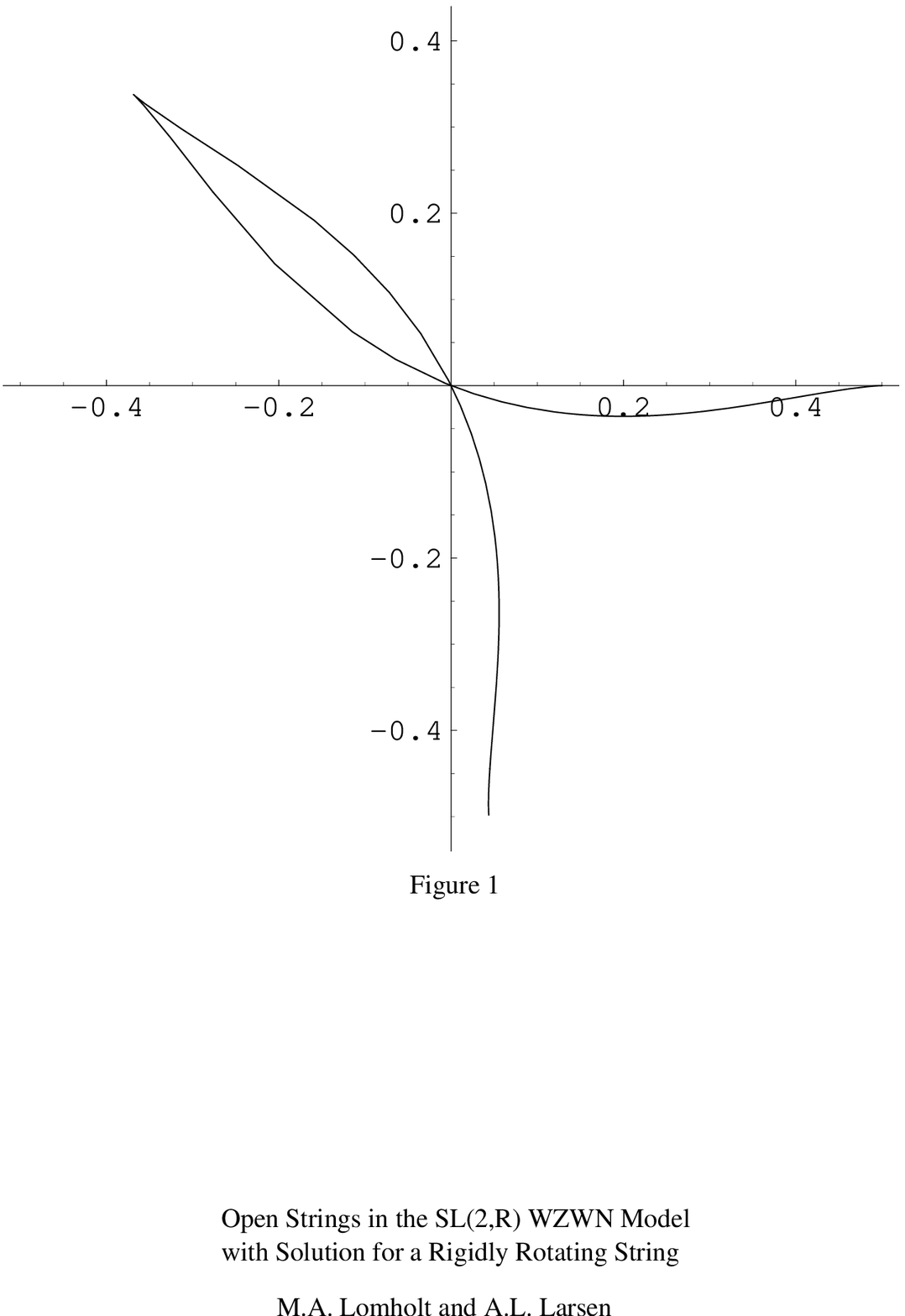}
\end{center}
\end{figure}


\newpage
\begin{thebibliography}{99}
\bibitem{nambu1}Y. Nambu, in {\it Symmetries and Quark models}, ed. R.
Chand (Gordon
and Breach, 1970).
\bibitem{nielsen}H.B. Nielsen,  submitted to the
15th
International Conference on High Energy Physics, Kiev, 1970.
\bibitem{susskind}L. Susskind, Nuovo Cim. A{\bf 69}, 457 (1970).
\bibitem{nambu2} Y. Nambu, {\it Duality and hydrodynamics}, Lectures at
the
Copenhagen symposium, 1970.
\bibitem{goto}T. Goto, Prog. Theor. Phys. {\bf 46}, 1560 (1971).
\bibitem{inigo}H.J. de Vega and I.L. Egusquiza, Phys. Rev. D{\bf 54}, 7513
(1996).
\bibitem{frolov}V. Frolov, S. Hendy and J.P. De Villiers, Class. Quant.
Grav. {\bf 14}, 1099 (1997).
\bibitem{kar}S. Kar and S. Mahapatra, Class. Quant. Grav. {\bf15}, 1421
(1998).
\bibitem{green}M.B. Green, J.H. Schwarz and E. Witten, {\it Superstring
Theory} (Cambridge University Press, 1987).
\bibitem{pol1}J. Polchinski, Phys. Rev. Lett. {\bf 75}, 4724 (1995).
\bibitem{pol2}J. Polchinski, {\it TASI Lectures on D-branes},
hep-th/9611050, 1996
(unpublished).
\bibitem{pol3}J. Polchinski, {\it String Theory} (Cambridge
University Press, 1998)
\bibitem{klimcik}C. Klimcik and P. Severa, Nucl. Phys. B{\bf 488}, 653
(1997).
\bibitem{kato}M. Kato and T. Okada, Nucl. Phys. B{\bf 499}, 583 (1997).
\bibitem{tseytlin}S. Stanciu and A. Tseytlin, JHEP {\bf 9806}, 010 (1998).
\bibitem{alekseev}A. Y. Alekseev and V. Schomerus, Phys. Rev. D{\bf 60},
061901 (1999).
\bibitem{stanciu}S. Stanciu, JHEP {\bf 9909}, 028 (1999).
\bibitem{bachas}C. Bachas and M. Petropoulos, JHEP {\bf 0102}, 025 (2001).
\bibitem{ribault}P.M. Petropoulos and S. Ribault, JHEP  {\bf 0107}, 036 (2001).
\bibitem{lee}P. Lee, H. Ooguri, J. Park and J. Tannenhauser, Nucl. Phys.
{\bf B610}
(2001).
\bibitem{bachas2}C. Bachas, {\it D-branes in some near-horizon
geometries},
hep-th/0106234, 2001 (unpublished).
\bibitem{witten}E. Witten, Commun. Math. Phys. {\bf 92}, 455 (1984).
\bibitem{balog}J. Balog, L. O'Raifeartaigh, P. Forgacs and A. Wipf, Nucl.
Phys. B{\bf 325}, 225 (1989).
\bibitem{dixon}L.J. Dixon, M.E. Peskin and J. Lykken, Nucl. Phys. B{\bf
325}, 329 (1989).
\bibitem{petr}P. Petropoulos, Phys. Lett. B{\bf 236}, 151 (1990).
\bibitem{moh}N. Mohammedi, Int. Journ. Mod. Phys. A{\bf 5}, 3201 (1990).
\bibitem{nem}I. Bars and D. Nemeschansky, Nucl. Phys. B{\bf 348},
89 (1991).
\bibitem{bars}I. Bars, Nucl. Phys. B{\bf 334}, 125 (1990).
\bibitem{hwa}S. Hwang, Nucl. Phys. B{\bf 354}, 100 (1991).
\bibitem{maans1}M. Henningson and S. Hwang, Phys. Lett. B{\bf 258},
341 (1991).
\bibitem{maans2}M. Henningson, S. Hwang, P. Roberts and B. Sundborg, Phys.
Lett. B{\bf 267}, 350 (1991).
\bibitem{hwang}S. Hwang, Phys. Lett. B{\bf 276}, 451 (1992).
\bibitem{malda}J. Maldacena, Adv. Theor. Math. Phys. {\bf 2}, 231 (1998).
\bibitem{bars1}I. Bars, Phys. Rev. D{\bf 53}, 3308 (1996).
\bibitem{bars2}I. Bars, in {\it String Gravity and
Physics at the Planck Energy Scale}, eds. N. S\'{a}nchez and A.
Zichichi (Kluwer Academic Publishers, 1995).
\bibitem{ooguri}J. Maldacena and H. Ooguri, J. Math. Phys. {\bf 42} 2929
(2001).

\end{thebibliography}
\end{document}